\documentclass[hyper]{JHEP}
\usepackage{graphicx,amssymb,bm,latexsym,amsmath}

\title{Pulsating strings on $(AdS_3 \times S^3)_\varkappa$ }
\author{Kamal L. Panigrahi, Pabitra M. Pradhan and Manoranjan Samal\\
Department of Physics,\\Indian Institute of Technology Kharagpur,\\
Kharagpur-721 302, India\\
Email: \email{panigrahi,ppabitra,manoranjan@phy.iitkgp.ernet.in}}

\abstract{We derive the energy of pulsating strings as a function
of adiabatic invariant oscillation number, which oscillates in
$S^2_{\varkappa}$. We find similar solutions for the strings
oscillating in deformed $AdS_3$. Furthermore, we generalize the
result of the oscillating strings in anti–de Sitter space in the
presence of extra angular momentum in $(AdS_3 \times
S^1)_\varkappa$.}


\begin{document}

\section{Introduction}
The conjectured duality between the supersymmetric Yang-Mills
theory in four dimensions and type IIB superstring in the
compactified AdS space \cite{Maldacena:1997re} has been the major
research area for recent few years. Though solving exact free
string spectrum on a generic given background is highly
non-trivial problem, robustness of integrability in the both side
of the conjecture played a key role in reducing the problem of
solving the spectra in the large charge limit to the the problem
of solving a set of algebraic Bethe equations. The fact that the
lagrangian field equations of the $AdS_5 \times S^5$ theory can be
recast in the zero curvature form \cite{Bena:2003wd} introduces
the integrability on the anti-de Sitter side of the correspondence
which ensures the existence of an infinite number of conserved
quantities. The integrability arises as a quantum symmetry of
operator mixing in CFT side \cite{Minahan:2002ve},
\cite{Beisert:2003yb} and as a classical symmetry on the string
world-sheet in AdS space \cite{Bena:2003wd}. Under the assumption
that integrability continues to hold at the quantum level, the
spectrum of the $AdS_5 \times S^5$ superstring is determined by
means of the thermodynamic Bethe ansatz applied to a doubly Wick
rotated version of its world sheet theory \cite{Zamo:1990ab},
\cite{Dorey:1996re}. Precisely, the integrability has improved the
understanding of the equivalence between the Bethe equation for
the spin chain and the corresponding classical realization of
Bethe equation for the classical $AdS_5 \times S^5$ string sigma
model \cite{Kazakov:2004qf}, \cite{Zarembo:2004hp}. The
corresponding Bethe equations are based on the knowledge of the
S-matrix which describes the scattering of world-sheet excitations
of the gauge-fixed string sigma model or the excitations of a
certain spin chain in the dual gauge theory \cite{Kazakov:2004qf},
\cite{Beisert:2004hm}, \cite{Arutyunov:2004vx},
\cite{Staudacher:2004tk}, \cite{Beisert:2005fw},
\cite{Beisert:2005tm}.

To improve our understanding of the relationship between
integrability and the amount of global symmetries preserved by the
target space-time, one should explore possibilities of various
deformations of the string target space time that preserve the
integrability of the two-dimensional quantum field theory on the
world sheet. Integrable deformations of $Ads_5 \times S^5$ can be
achieved by a combination of T- duality and shift transformations
\cite{Lunin:2005jy}, \cite{Frolov:2005dj}. This geometric approach
results in a new class of deformations which can be described in
terms of original string theory and the deformations result into
quasi-periodic but keeping the integrability intact. The other way
is an algebraic approach based on q-deformations of the world
sheet S-matrix \cite{Beisert:2008tw}, \cite{Beisert:2012wt},
\cite{Hoare:2011wr}, \cite{deLeeuw:2011fr}, \cite{Hoare:2012fc},
\cite{Arutyunov:2012zt}, \cite{Arutyunov:2012ai},
\cite{Hoare:2013ysa}. Recently one real deformed parametered
integrable q-deformed $AdS_5 \times S^5$ super coset model with
fermionic degree of freedom was found in \cite{Delduc:2013qra}.
The deformed background breaks the symmetry of $AdS_5 \times S^5$
to $[U(1)]^6$, which urges to the new insight of its dual field
theory which has to be explored yet. In order to understand the
various aspects of the background one can look in to
\cite{Arutyunov:2013ega}, \cite{Hoare:2014pna},
\cite{Arutynov:2014ota}, \cite{Kameyama:2014bua},
\cite{Khouchen:2014kaa}, \cite{Ahn:2014aqa},
\cite{Arutyunov:2014cra}, \cite{Arutyunov:2014cda},
\cite{Banerjee:2014bca}. The perturbative world sheet scattering
matrix of bosonic particles of the model was computed in
\cite{Arutyunov:2013ega}. The maximal deformation limit of this
model is T-dual to a flipped double Wick rotation of the target
space and in the imaginary limit it becomes that of a pp-wave
background with a curved transverse part \cite{Hoare:2014pna}.
Thermodynamic Bethe Ansatz description of exact finite size
spectra concludes that this model maps on to itself under double
Wick rotation \cite{Arutynov:2014ota}. The classical integrable
structure of anisotropic Landau-Lifshitz sigma models has been
derived by taking fast moving string limits in the bosonic sub
sector of this model \cite{Kameyama:2014bua}. This background is
formally related to $dS_5 \times H^5$ by a double T-duality with
hidden supersymmetry \cite{Arutyunov:2014cra}. The bosonic
spinning strings on this background can be viewed as the solution
to a deformed Neumann model \cite{Arutyunov:2014cda}. In this
deformed supercoset model corresponding type IIB supergravity
solutions in the subset of $AdS_2 \times S^2$ and $AdS_3 \times
S^3$ have been computed with non-trivial dilaton and RR scalar
with a free parameter dependency on the solution
\cite{Lunin:2014tsa}. Further by following the Yang-Baxter sigma
model approach with classical $r$-matrices which satisfy the
classical Yang-Baxter equation and  carry two parameters and
three-parameter generalization, type IIB supergravity solutuions
have been found in \cite{Matsumoto:2014ubv}. However, the
existence and properties of a gauge theory dual to string theory
in the deformed background is still an open question. In this
connection giant magnons and its finite size correction
\cite{Khouchen:2014kaa}, \cite{Ahn:2014aqa},
\cite{Banerjee:2014bca} have been computed for rotating string in
string theory side. Here we wish to study pulsating string
solution in the sub sectors of the deformed background as they are
more stable than rotating ones \cite{Khan:2005fc}. After the
inception of the pulsating string in \cite{Gubser:2002tv}, they
have been studied both in AdS and non-AdS background
\cite{Minahan:2002rc}, \cite{Khan:2003sm}, \cite{Engquist:2003rn},
\cite{Arutyunov:2003za}, \cite{Dimov:2004xi},
\cite{Smedback:1998yn}, \cite{Kruczenski:2004cn},
\cite{Bobev:2004id}, \cite{Park:2005kt}, \cite{deVega:1994yz},
\cite{Chen:2008qq}, \cite{Dimov:2009rd}, \cite{Arnaudov:2010by},
\cite{Arnaudov:2010dk}, \cite{Beccaria:2010zn},
\cite{Giardino:2011jy}, \cite{Pradhan:2013sja},
\cite{Pradhan:2014zqa}.

The rest of the paper is organized as follows. in section-2, we
preview the truncated models of $\varkappa$-deformed $AdS_5 \times
S^5$. In section-3, we study the semiclassical oscillating string
solution in the deformed $R \times S^2$. In section-4, we analyze
the solution in terms of energy as function of oscillation number
for a class of pulsating strings in the deformed $AdS_3$. In
section-5, we generalize the previous section with an extra
angular momentum in the $S^1$. In section-6, we conclude with some
remarks.
\section{Consistent truncations of $(AdS_5 \times S^5)_\varkappa$}
As $(AdS_5 \times S^5)_\varkappa$ is a classically integrable
background, its consistent truncations must be classically
integrable. Truncated lower dimensional integrable string models
have been computed in \cite{Hoare:2014pna}. We write the relevant
backgrounds here.
\begin{eqnarray}
ds^2_{AdS_3 \times S^3} = - h(\rho) dt^2 + f(\rho) d\rho^2 +
\rho^2 d\phi^2 + {\tilde h}(r) d\varphi^2 + \tilde{f} (r) dr^2 +
r^2 d\psi^2 , \label{1}
\end{eqnarray}
where
\begin{eqnarray}
h(\rho) = \frac{1+\rho^2}{1-\varkappa^2 \rho^2}, ~~~ f(\rho) =
\frac{1}{(1+\rho^2)(1-\varkappa^2 \rho^2)} \nonumber \\
{\tilde h}(r) = \frac{1-r^2}{1+\varkappa^2 r^2}, ~~~~ {\tilde
f}(r) = \frac{1}{(1-r^2)(1+\varkappa^2 r^2)}. \nonumber
\end{eqnarray}
With $\phi = \psi = 0$, we get the lower dimensional consistent
background as
\begin{eqnarray}
ds^2_{AdS_2 \times S^2} = - h(\rho) dt^2 + f(\rho) d\rho^2 +
{\tilde h}(r) d\varphi^2 + \tilde{f} (r) dr^2. \label{2}
\end{eqnarray}
Here co-ordinates have their usual range as in case of undeformed
one and $\varkappa \in [0,\infty).$

\section{Pulsating string in $S_{\varkappa}^2$}
Here we wish to study the string solution to a class of pulsating
string which is oscillating in deformed $S^2$. In order to get the
metric, we substitute the followings in the equation (\ref{2})
\begin{equation}
\rho = 0, ~~~~~ \varphi = \phi, ~~~~~ r = \cos\psi, \label{3}
\end{equation}
and get
\begin{eqnarray}
ds^2 = -dt^2 + \frac{d\psi ^2}{1+\varkappa ^2 \cos^2 \psi} +
\frac{\sin^2 \psi d\phi ^2 }{1+\varkappa ^2 \cos ^2 \psi }.
\label{4}
\end{eqnarray}
The Polyakov action of the metric is given by
\begin{eqnarray}
I=\frac{\sqrt{\hat{\lambda}}}{4 \pi}\int d\tau \sigma
\left[-(\dot{t}^2 - {t^ \prime}^2) + \frac{\dot{\psi}^2 - {\psi ^
\prime}^2}{1+ \varkappa^2 \cos^2 \psi }+ \frac{ \dot{ \phi} ^2 -
{\phi ^ \prime }^2}{1+\varkappa ^2 \cos^2 \psi} \sin^2\psi \right
], \label{5}
\end{eqnarray}
where the `dot' and `prime' denote the derivatives with respect to
$\tau$ and $\sigma$ respectively and $ \hat\lambda = \lambda
(1+\varkappa^2)$, where $\lambda$ is the `t Hooft coupling
constant. We write the following anstaz for the pulsating string
\begin{equation}
t = \kappa \tau ,~~~~~ \psi = \psi(\tau), ~~~~~  \phi = m \sigma.
\label{6}
\end{equation}
Equation of motion for $\psi $ is given by
\begin{eqnarray}
(1+\varkappa ^2 \cos^2 \psi)\left(\frac{2 \ddot{\psi}}{\sin 2
\psi}+ m^2 \right) +\varkappa ^2 \dot{\psi} ^2 +\varkappa ^2 m^2
\sin ^2 \psi = 0. \label{7}
\end{eqnarray}
From the Virassoro constraint we get
\begin{eqnarray}
\frac{m^2 \sin^2 \psi}{1+ \varkappa ^2 \cos ^2 \psi} - \kappa^2 +
\frac{\dot{\psi}^2 }{1+ \varkappa ^2 \cos ^2 \psi } = 0. \label{8}
\end{eqnarray}
The energy for this string configuration is given by
\begin{eqnarray}
E = \sqrt{\hat{\lambda}}~\varepsilon = \sqrt{\hat{\lambda}}~
\kappa. \label{9}
\end{eqnarray}
The canonical momentum associated with $\psi$ is
\begin{equation}
\Pi_\psi = \frac{\dot\psi}{1+ \varkappa^2 \cos^2 \psi}~~.
\label{10}
\end{equation}
From Equation(\ref{8}) we get
\begin{eqnarray}
\dot{\psi}^2 =  \varepsilon ^2 ( 1+ \varkappa ^2 \cos^2 \psi) -
m^2 \sin ^2 \psi~. \label{11}
\end{eqnarray}
We can compute the oscillation number which should take integer
values in quantum theory as
\begin{eqnarray}
N &=& \sqrt{\hat{\lambda}}~ \mathcal{N} =
\frac{\sqrt{\hat{\lambda}}} {2\pi} \oint d\psi~ \Pi_\psi \cr & \cr
&=&  \frac{\sqrt{\hat{\lambda}}}{2\pi} \oint d\psi
\sqrt{\frac{\varepsilon^2 }{1+ \varkappa ^2  \cos  ^2 \psi } -
\frac{m^2 \sin ^2 \psi}{(1+ \varkappa ^2  \cos ^2 \psi)^2 }}~.
\label{12}
\end{eqnarray}
Substituting $ \sin ^2 \psi  = z $ in the above equation
(\ref{12}) we get,
\begin{eqnarray}
\mathcal{N}= \frac{1}{2\pi} \int_ 0 ^ b \frac{dz}{1+ \varkappa ^2
- \varkappa ^2 z} ~ \sqrt{\frac{\varepsilon ^2(1+\varkappa ^ 2 -
\varkappa ^2 z)- m^2 z}{(1-z)z}}~. \label{13}
\end{eqnarray}
To find out this integration ,we have taken derivative of $N $
with respect to $m$, i.e
\begin{eqnarray}
\frac{\partial \mathcal{ N}}{\partial m}= - \frac{m}{\pi}  \int_ 0
^ b \frac{z~dz}{(1+ \varkappa ^2 - \varkappa ^2 z) \sqrt{z
\varepsilon ^2 (1-z)(1+\varkappa ^ 2 - \varkappa ^2 z)- m^2
z^2(1-z)}}, \label{14}
\end{eqnarray}
where $a>b>c$ are roots of the polynomial
\begin{equation}
f(z)=z \varepsilon ^2 (1-z)(1+\varkappa ^ 2 - \varkappa ^2 z)- m^2
z^2(1-z).
\end{equation}
And $~~ a=1,~~~~~~b = \frac{\varepsilon ^2 (1+\varkappa ^2)}{m^2 +
\varepsilon ^2 \varkappa ^2 },~~~~~~c=0, ~~~~~~ d=
\frac{1+\varkappa^2}
{\varkappa^2} $.\\
The above integral in (\ref{14}) can be written as sum of two
integrals i.e.
\begin{eqnarray}
\frac{\partial \mathcal{ N}}{\partial m} = I_1 + I_2 ~,  \nonumber
\end{eqnarray}
Where
\begin{eqnarray}
I_1 &=& \frac{m}{ \pi \varkappa ^2 }\int _0 ^ b \frac{dz}{\sqrt{z
\varepsilon ^2 (1-z)(1+\varkappa ^ 2 - \varkappa ^2 z)- m^2
z^2(1-z)}} \nonumber \\
&=&  \frac{m}{ \pi \varkappa ^2 \sqrt{m^2 +\varepsilon ^2
\varkappa ^2 }}
\int _0 ^ b \frac{ dz}{\sqrt{(z-a)(z-b)(z-c)}} \nonumber \\
&=& \frac{m}{\varkappa ^2 \pi \sqrt{m^2 +\varepsilon ^2 \varkappa
^2 }}~ \mathbb{K}\left[ b \right],
\end{eqnarray}
and
\begin{eqnarray}
I_2 &=& \frac{m}{ \pi \varkappa ^2 } \int _0 ^ b \frac{-d ~
dz}{(d-z)\sqrt{z \varepsilon ^2 (1-z)(1+\varkappa ^ 2 - \varkappa
^2 z)- m^2
z^2(1-z)} } \nonumber \\
&=& \frac{-m}{ \pi \varkappa ^2 \sqrt{m^2 +\varepsilon ^2 \varkappa ^2 }}
\int _0 ^ b \frac{d~dz}{(d-z)\sqrt{(z-a)(z-b)(z-c)}} \nonumber \\
&=& \frac{-m}{\varkappa ^2 \pi \sqrt{m^2 +\varepsilon ^2 \varkappa
^2 }} ~\Pi \left [ \frac{b}{r},b \right ].
\end{eqnarray}
Now
\begin{eqnarray}
\frac{\partial \mathcal{ N}}{\partial m}=\frac{m}{\varkappa ^2
\pi \sqrt{m^2 +\varepsilon ^2 \varkappa ^2 }} \left
(\mathbb{K}\left[ b \right]- \Pi \left [  \frac{b}{r},b \right ]
\right ), \label{15}
\end{eqnarray}
where $ \mathbb{K} $, $\Pi $ are  complete elliptical integral of first and third kind respectively and
Expanding the equation (\ref{15}) for small value $\varepsilon$ in
the short string limit
\begin{eqnarray}
\frac{\partial \mathcal{ N}}{\partial m} = \frac{- \varepsilon
^2}{2m ^2} + \frac{3(-1+ \varkappa ^2)}{16 m^4 } \varepsilon ^4
- \frac{5(3-2 \varkappa 2 + 3\varkappa 4)}{128 m^6} \varepsilon ^6\mathcal{O}[\varepsilon ^8]. \label{17}
\end{eqnarray}
Taking integration with respect to $m$ we get
\begin{eqnarray}
\mathcal{N}= \frac{\varepsilon ^2}{2 m} + \frac{1- \varkappa ^2
}{16 m^3 } \varepsilon ^4  +\frac{3-2\varkappa ^2 + 3 \varkappa ^4}{128 m ^5} \varepsilon ^6+ \mathcal {O}[\varepsilon ^ 8].
\label{18}
\end{eqnarray}
Reversing the series we get
\begin{eqnarray}
\varepsilon=  \sqrt{2m\mathcal{N}}\left(1- \frac{1-\varkappa
^2}{8m}\mathcal{N}-\frac{5+6 \varkappa ^2 + 5 \varkappa ^4}{ 128 m
^2}\mathcal{N}^2 +\mathcal{O} [\mathcal{N}^3] \right). \label{19}
\end{eqnarray}
In the above dispersion relation $\varepsilon < m$ gives an upper
bound for $\mathcal{N}$, so one cannot take the large
$\mathcal{N}$ limit. This gives the short string or small
oscillation number expansion of the classical energy. If we put
$\varkappa \rightarrow 0$ in the above equation (\ref{19}), we get
the exact expression for undeformed $S^2$ as found in
\cite{Beccaria:2010zn}.

\section{Pulsating string in deformed $AdS_3$}
In this section we study the semiclassical quantization of a class
of strings which is oscillating in the radial $\rho$ direction of
$AdS_3$. We get the relevant metric for this from
equation(\ref{1}) (taking only $AdS$ part) with the substitution
of $\rho = \sinh\rho$
\begin{eqnarray}
ds^2 = - \frac{ \cosh^2\rho }{1-\varkappa^2\sinh^2\rho}dt^2 +
\frac{d\rho^2}{1-\varkappa^2 \sinh^2\rho} + \sinh^2\rho d\phi^2.
\label{20}
\end{eqnarray}
We chose the ansatz for this configuration as
\begin{eqnarray}
t = t(\tau) ,~~~~~ \rho = \rho(\tau), ~~~~~  \phi &=& m \sigma.
\label{21}
\end{eqnarray}
The polyakov action of the given metric is given by
\begin{eqnarray}
I = \frac{\sqrt{\hat{\lambda}}}{4 \pi} \int d\tau d\sigma
\left[-~\frac{ \cosh^2\rho }{1-\varkappa^2\sinh^2\rho}\dot{t}^2 +
\frac{\dot{\rho}^2}{1-\varkappa ^2 \sinh^2\rho} + m^2
\sinh^2\rho\right]. \label{22}
\end{eqnarray}
Equation of motion for t and $\rho$  are given by
\begin{eqnarray}
\ddot{t}\cosh^2\rho + \dot{\rho}\dot{t} \sinh 2\rho [1+
\frac{\varkappa^2\cosh^2\rho}{1-\varkappa^2\sinh^2\rho}] &=& 0
\\
2 \ddot{\rho}(1-\varkappa ^2 \sinh ^2 \rho)+ \sinh \rho [m^2
(1-\varkappa ^2 \sinh^2 \rho)^2 + \varkappa ^2 \dot{\rho} ^2 +
\dot{t} ^2 (1+ \varkappa ^2)] &=& 0. \label{23}
\end{eqnarray}
The Virasoro constraint gives us
\begin{eqnarray}
m^2\sinh^2\rho - \frac{\cosh^2\rho}{1-\varkappa ^2
\sinh^2\rho}\dot{t}^2+\frac{1}{1-\varkappa
^2\sinh^2\rho}\dot{\rho}^2 = 0. \label{24}
\end{eqnarray}
The energy of the oscillating string is given by
\begin{eqnarray}
E = \sqrt{\hat{\lambda}} \varepsilon = \frac{\cosh^2\rho}
{1-\varkappa^2 \sinh^2\rho} \dot{t}. \label{25}
\end{eqnarray}
The canonical momentum associated with $\rho$ is
\begin{equation}
\Pi_{\rho} = \frac{\dot{\rho}} {1-\varkappa^2 \sinh^2\rho}.
\label{26}
\end{equation}
Using the equations (\ref{24}) and (\ref{25}), we can get
\begin{eqnarray}
\dot{\rho}^2 - \frac{\varepsilon ^2 (1-\varkappa^2 \sinh^2\rho)^2}
{\cosh^2 \rho} + m^2\sinh^2\rho (1-\varkappa^2 \sinh^2\rho) = 0.
\label{27}
\end{eqnarray}
With the help of equation (\ref{26}), we can write
\begin{equation}
\Pi^2_{\rho} + V(\rho) = 0, ~~~~~~ V(\rho) = -
\frac{\varepsilon^2} {\cosh^2\rho} + \frac{m^2\sinh^2\rho}
{1-\varkappa^2\sinh^2\rho}. \label{28}
\end{equation}
This may be interpreted as an equation for a particle moving in a
potential which is growing to infinity at $\rho \rightarrow
\infty$. The coordinate $\rho(\tau)$ thus oscillates between $0$
and a maximal $\rho$ value $(\rho_{max})$. Since the string is
oscillating along $\rho$ direction ,we can define the oscillation
number as
\begin{eqnarray}
N &=& \sqrt{\hat{\lambda}} ~\mathcal{N} =
\frac{\sqrt{\hat{\lambda}}} {2\pi} \oint d\rho ~ \Pi _\rho \cr &
\cr &=& \frac{\sqrt{\hat{\lambda}}} {\pi} \int_0^{\rho_{max}}
\sqrt{\frac{\varepsilon^2}{\cosh^2\rho} - \frac{m^2\sinh^2\rho}
{1-\varkappa^2\sinh^2\rho}} ~. \label{29}
\end{eqnarray}
Taking $\sinh ^2 \rho = z$
\begin{eqnarray}
\mathcal{N} &=& \frac{1}{2\pi} \int _0^ {R_2}
\frac{dz}{1+z}\sqrt{\frac{\varepsilon^2(1-\varkappa^2 z)-m^2
z(1+z)}{z(1-\varkappa^2 z)}}. \label{30}
\end{eqnarray}
To make the integration simple we make the derivative with respect to m
\begin{eqnarray}
\frac{\partial{\mathcal{N}}}{\partial{m}} = - \frac{m}{2 \pi}
\int_0^{R_2} dz \frac{\sqrt{z}}{\sqrt{\varepsilon^2(1-\varkappa^2
z)^2 -m^2 z(1+z)(1-\varkappa^2 z)}}, \label{31}
\end{eqnarray}
where $R_1>R_2>R_3$ are roots of the polynomial
\begin{equation}
f(z) = \varepsilon^2(1-\varkappa^2 z)^2 -m^2 z(1+z)(1-\varkappa^2
z).
\end{equation}
And
\begin{eqnarray}
R_1 = \frac{1}{\varkappa ^ 2},~~~~ R_2 &=& \frac{-m^2 -\varepsilon
^2 \varkappa ^2 + \sqrt{4 m^2 \varepsilon ^2 + (m^2 + \varepsilon
^2 \varkappa ^2 )^2 }}{2m^2}, \nonumber \\ R_3 &=& \frac{-m^2
-\varepsilon ^2 \varkappa ^2 - \sqrt{4 m^2 \varepsilon ^2 + (m^2 +
\varepsilon ^2 \varkappa ^2 )^2 }}{2m^2}.\nonumber
\end{eqnarray}
The above integral can be written in the standard elliptical
integrals as
\begin{eqnarray}
\frac{\partial{\mathcal{N}}}{\partial{m}} = \frac{R_3}{\pi
\sqrt{R_1(R_2 - R_3)}\varkappa} \left[\Pi\left(\frac{R_2}{R_2 -
R_3} ,\frac{R_2(R_1 -R_3)}{R_1(R_2 -R_3)}\right)
-\mathbb{K}\left(\frac{R_2(R_1 -R_3)}{R_1(R_2
-R_3)}\right)\right], \nonumber \\ \label{32}
\end{eqnarray}
Now expanding the above equation for a small oscillation number
with small $\varepsilon $ we will get
\begin{eqnarray}
\frac{\partial{\mathcal{N}}}{\partial{m}} &=& \frac{-\varepsilon
^2 }{4 m^2 } +\frac{3(5+3 \varkappa ^2)}{32 m^4} \varepsilon ^4 -
\frac{5(63+70 \varkappa ^2 +15 \varkappa ^4)}{256 m^5} \varepsilon
^6 + \mathcal{O}\left[\varepsilon^8 \right]. \label{33}
\end{eqnarray}
Integrating with respect to m we get
\begin{eqnarray}
\mathcal{N} = \frac{\varepsilon ^2 }{4 m } -\frac{(5+3 \varkappa
^2)}{32 m^3} \varepsilon ^4 + \frac{(63+70 \varkappa ^2 +15
\varkappa ^4)}{256 m^5} \varepsilon ^6 + \mathcal{O}
\left[\varepsilon^8 \right]. \label{34}
\end{eqnarray}
Reversing the series
\begin{eqnarray}
\varepsilon = 2 \sqrt{m \mathcal{N}} +\frac{5+3 \varkappa ^2}
{2\sqrt{m}} \mathcal{N}^{3/2} +\frac{(-77 -70 +3 \varkappa ^4
)}{16 m^ {3/2}} \mathcal{N} ^{5/2} + \mathcal{O}
\left[\mathcal{N}^{7/2}\right]. \label{35}
\end{eqnarray}
This is the classical energy expression in the small energy limit
for short string configuration in deformed $AdS_3$. After
substituting $\varkappa = 0$, we can get the the flat-space
dependence which is expected in the small-energy limit where the
string oscillates near the center of $AdS_3$ which can be found in
\cite{Park:2005kt}. The result found in \cite{Beccaria:2010zn}
differs by a factor 2 as they have defined the oscillation number
accordingly.

\section{Pulsating  string in $(AdS_3 \times S^1)_\varkappa$}
In this section we generalize the previous section where we study
a class of oscillating string solution which is oscillating in the
radial $\rho$ direction of $AdS_3$ with an extra angular momentum
along $S^1$. In order to get the consistent truncated metric, we
substitute the following in the equation (\ref{1})
\begin{equation}
\rho = \sinh\rho, ~~~~~~~ r = \psi = 0. \label{36}
\end{equation}
Now the relevant background is given by
\begin{equation}
ds^2 = - \frac{ \cosh^2\rho }{1-\varkappa^2\sinh^2\rho}dt^2 +
\frac{d\rho^2}{1-\varkappa^2 \sinh^2\rho} + \sinh^2\rho d\phi^2 +
d\varphi ^2 . \label{37}
\end{equation}
Choosing the ansatz as
\begin{eqnarray}
t = t(\tau),~~~~ \rho = \rho(\tau),~~~~ \phi = m\sigma ,~~~~
\varphi =\varphi(\tau), \label{38}
\end{eqnarray}
we write down the Polyakov action of the above metric
\begin{equation}
I = \frac{\sqrt{\hat{\lambda}}} {4 \pi} \int~ d\tau d\sigma
\left[- \frac{ \cosh^2\rho } {1-\varkappa^2\sinh^2\rho} \dot{t}^2
+ \frac{\dot{\rho}^2} {1-\varkappa ^2 \sinh^2\rho} + m^2
\sinh^2\rho + \dot{\varphi} ^2 \right]. \label{39}
\end{equation}
Equation of motion for t is
\begin{equation}
\ddot{t} \cosh^2 \rho + 2 \dot{\rho} \dot{t} \sinh\rho \cosh \rho
\left[ 1+ \frac{\varkappa ^2 \cosh^2 \rho}{1-\varkappa ^2 \sinh^2
\rho} \right] = 0. \label{40}
\end{equation}
Equation of motion for $ \rho $ is
\begin{equation}
- 2 \ddot{\rho}(1-\varkappa ^2 \sinh ^2 \rho)+ \sinh \rho [m^2
(1-\varkappa ^2 \sinh^2 \rho)^2 + \varkappa ^2 \dot{\rho} ^2 +
\dot{t} ^2 (1+ \varkappa ^2)] = 0. \label{41}
\end{equation}
The Virassoro constraint is given by
\begin{eqnarray}
m^2\sinh^2\rho - \frac{\cosh^2\rho}{1-\varkappa ^2
\sinh^2\rho}\dot{t}^2+\frac{1}{1-\varkappa
^2\sinh^2\rho}\dot{\rho}^2 + \dot{\varphi} ^2 = 0. \label{42}
\end{eqnarray}
Conserved quantities are
\begin{eqnarray}
E &=& \sqrt{\hat{\lambda}} \varepsilon = \sqrt{\hat{\lambda}}
\frac{\cosh^2\rho} {1-\varkappa^2 \sinh^2\rho}\dot{t} \cr & \cr J
&=& \sqrt{\hat{\lambda}} j = \sqrt{\hat{\lambda}} \dot{\varphi}.
\label{43}
\end{eqnarray}
The canonical momentum associated with $\rho$ is
\begin{equation}
\Pi_\rho = \frac{\dot{\rho}} {1-\varkappa^2 \sinh^2\rho}.
\label{44}
\end{equation}
From the equation (\ref{42}), with the help of the equation
(\ref{43}) we get
\begin{equation}
\dot{\rho}^2 = \frac{\varepsilon ^2 (1-\varkappa^2 \sinh^2\rho)^2}
{\cosh^2 \rho} - (m^2\sinh^2\rho + j^2) (1-\varkappa^2
\sinh^2\rho). \label{45}
\end{equation}
With the help of the equation (\ref{44}), the above equation can
be written as
\begin{equation}
\Pi^2_{\rho} + V(\rho) = 0, ~~~~~~ V(\rho) = -
\frac{\varepsilon^2} {\cosh^2\rho} + \frac{m^2\sinh^2\rho}
{1-\varkappa^2\sinh^2\rho} + \frac{j^2} {1-\varkappa^2 \sinh^2
\rho}. \label{46}
\end{equation}
This is similar to the previous section (eq \ref{28}) with an
extra additive term. This is similar to an equation for a particle
moving in such a potential so that the coordinate $\rho(\tau)$
oscillates between $0$ and a maximal $\rho$ value $(\rho_{max})$.
Now we can write the oscillation number as
\begin{eqnarray}
N &=& \sqrt{\hat{\lambda}} ~\mathcal{N} =
\frac{\sqrt{\hat{\lambda}}} {2\pi} \oint d\rho ~ \Pi _\rho \cr &
\cr &=& \frac{\sqrt{\hat{\lambda}}} {2\pi}\oint d\rho
\sqrt{\frac{\varepsilon^2} {\cosh^2\rho} -\frac{m^2\sinh^2\rho}
{1-\varkappa^2\sinh^2\rho} - \frac{j^2} {1-\varkappa ^2 \sinh^2
\rho}}. \label{47}
\end{eqnarray}
Taking  $\sinh^2\rho = z $, then differentiating with respect to m
\begin{eqnarray}
\frac{\partial \mathcal{N}}{\partial m} &=& -~ \frac{m}{2\pi}
\int_0 ^ {R_2} dz \frac{ \sqrt{z}}{\sqrt{\varepsilon ^2
(1-\varkappa ^2 z)^2 - (m^2 z + j^2) (1-\varkappa ^2 z)(1+z)}},
\label{48}
\end{eqnarray}
where $R_1 > R_2 > R_3$ are roots of the polynomial
\begin{equation}
f(z) =\varepsilon ^2 (1-\varkappa ^2 z)^2 - (m^2 z + j^2)
(1-\varkappa ^2 z)(1+z).
\end{equation}
And
\begin{eqnarray}
R_1= \frac{1}{\varkappa ^ 2},~~~R_2 &=& \frac{-j^2 -m^2
-\varepsilon ^2 \varkappa ^2 + \sqrt{4 m^2 (\varepsilon ^2 - j^2)
+ (j^2+m^2 +
\varepsilon ^2 \varkappa ^2 )^2 }}{2m^2}, \nonumber \\
R_3 &=& \frac{-j^2-m^2 -\varepsilon ^2  \varkappa ^2 - \sqrt{4
m^2( \varepsilon ^2 -j^2) + (j^2+m^2 + \varepsilon ^2 \varkappa ^2
)^2 }} {2m^2}. \nonumber
\end{eqnarray}
The above integral in (\ref{48}) can be written as
\begin{eqnarray}
\frac{\partial{\mathcal{N}}} {\partial{m}} = \frac{R_3} {\pi
\sqrt{R_1(R_2 - R_3)}\varkappa} \left[\Pi\left(\frac{R_2} {R_2 -
R_3} ,\frac{R_2(R_1 -R_3)} {R_1(R_2 -R_3)}\right) -\mathbb{K}
\left(\frac{R_2(R_1 -R_3)} {R_1(R_2 -R_3)}\right)
\right],\nonumber \\ \label{49}
\end{eqnarray}
$\mathbb{K} $ and $\Pi$ are complete elliptical integral of
first and third kind respectively. Expanding the above equation
with small $\varepsilon $ and small j
\begin{eqnarray}
&&\frac{\partial{\mathcal{N}}}{\partial{m}} = \left[ \frac{j^2}{4
m^2} +\frac{3 (1- \varkappa ^2 )}{32 m ^2 } j^4 + \mathcal{O}
[j^6] \right ] \nonumber \\  &&+ \left [- \frac{1}{4 m^2} -
\frac{3(3+\varkappa ^2)} {16 m^4 } j^2  - \frac{15(15+ 6 \varkappa
^2 - \varkappa ^4 )}
{256 m^6}j^4 + \mathcal{O} [j^6] \right] \varepsilon ^2  \nonumber \\
&&+ \left[ \frac{3(5+3\varkappa ^2 )}{32 m^4} + \frac{15(35+30
\varkappa ^2 + 3 \varkappa ^4)}{256 m^6} j^2 +
\frac{105(105+105\varkappa ^2 +15 \varkappa ^4- \varkappa ^6)}
{2048 m^8} j^4 + \mathcal{O} [j^6] \right ] \varepsilon^4 \nonumber \\
&&+ \mathcal{O} [\varepsilon^6] . \label{50}
\end{eqnarray}
Integrating with respect to m and reversing the series we get
\begin{eqnarray}
\varepsilon = 2 \sqrt{m \mathcal{M}} ~K_1(j) \left[1+ K_2(j)
\frac{5\mathcal{M}}{4m}+  \mathcal{O} [\mathcal{M} ^2] \right],
\label{51}
\end{eqnarray}
where
\begin{eqnarray}
\mathcal{M} &=& \mathcal{N} + \frac{j^2}{4m}+
\frac{j^4(1-\varkappa ^2)}{32 m^3} + \mathcal{O} [j^6], \nonumber \\
K_1(j) &=& 1+ \frac{3+ \varkappa ^2}{4 m^2} j^2+ \frac{3(15+ 6
\varkappa ^2 - \varkappa ^4)}{64 m^4} j^4 + \mathcal{O} [j^6],
\nonumber \\ K_2(j) &=& \frac{ 5+3\varkappa ^2}{5} + \frac{3(35+
30 \varkappa ^2  + 3 \varkappa ^4)}{40 m^2 } j^2+ \frac{3(105+105
\varkappa ^2 + 15 \varkappa ^4 - \varkappa ^6)}{64 m^4}j ^4 + \mathcal{O} [j^6]
\nonumber\\
\end{eqnarray}
This is the classical energy expression for the small energy and
angular momentum in the $\varkappa$-deformed $AdS_3 \times S^1$.
After putting $\varkappa = 0$, we can get the energy for the short
string which oscillates near the center of $AdS_3$ with an angular
momentum in $S^1$ in undeformed $AdS_3 \times S^1$ as computed in
\cite{Pradhan:2013sja}. With both $\varkappa$ and angular momentum
as zero we can get back the energy expression for the strings
oscillating in one plane for small energy limit as in the
\cite{Park:2005kt}.
\section{Conclusion}
We have studied various pulsating string in the so called
$\varkappa$ deformed $AdS_3 \times S^3$ background. We find the
energy of the short string in the small energy limit for the
pulsating strings in the $\varkappa$-deformed $S^3_{\varkappa}$
subspace of the full $(AdS_5 \times S^5)_{\varkappa}$ background.
$\varkappa = 0$ limit agrees with the computation of the
undeformed case and the $\varkappa$ infact enters in a vary
natural way in the expression. It is perhaps along the expected
lines as a theory with non-zero $\varkappa$ also provides an exact
integrable sigma model background (with a redefined string
tension) and hence the string configurations in the undeformed
background must also have correspondence with the ones in the
deformed case as well. We have further found out the short string
energy as a function of ${\cal N}, m, \varkappa$. We have also
analyzed case for the string with an extra angular momentum along
the deformed $S^1$. We wish to look for the field theory duals in
these theories in future.

\end{document}